\documentclass[conference]{IEEEtran}
\IEEEoverridecommandlockouts

\usepackage{url}

\usepackage{cite}
\usepackage{amsmath,amssymb,amsfonts}
\usepackage{algorithmic}
\usepackage{algorithm2e}
\usepackage{graphicx}
\usepackage{textcomp}
\usepackage{xcolor}
\usepackage{booktabs}

\begin{document}

\title{Static Analysis as a Feedback Loop:\\Enhancing LLM-Generated Code Beyond Correctness}

\author{\IEEEauthorblockN{Scott Blyth}
\IEEEauthorblockA{\textit{Monash University} \\
Melbourne, Australia \\
scottblyth202@gmail.com}
\and
\IEEEauthorblockN{Sherlock A. Licorish}
\IEEEauthorblockA{\textit{University of Otago} \\
Dunedin, New Zealand \\
sherlock.licorish@otago.ac.nz}
\and
\IEEEauthorblockN{Christoph Treude}
\IEEEauthorblockA{\textit{Singapore Management University} \\
Singapore, Singapore \\
ctreude@smu.edu.sg}
\and
\IEEEauthorblockN{Markus Wagner}
\IEEEauthorblockA{\textit{Monash University} \\
Melbourne, Australia\\
markus.wagner@monash.edu}
}

\thispagestyle{plain}\pagestyle{plain}

\maketitle

\begin{abstract}

Large language models (LLMs) have demonstrated impressive capabilities in code generation, achieving high scores on benchmarks such as HumanEval and MBPP. However, these benchmarks primarily assess functional correctness and neglect broader dimensions of code quality, including security, reliability, readability, and maintainability. In this work, we systematically evaluate the ability of LLMs to generate high-quality code across multiple dimensions using the PythonSecurityEval benchmark. We introduce an iterative static analysis-driven prompting algorithm that leverages Bandit and Pylint to identify and resolve code quality issues. Our experiments with GPT-4o show substantial improvements: security issues reduced from \textgreater40\% to 13\%, readability violations from \textgreater80\% to 11\%, and reliability warnings from \textgreater50\% to 11\% within ten iterations. These results demonstrate that LLMs, when guided by static analysis feedback, can significantly enhance code quality beyond functional correctness.

\end{abstract}



\section{Introduction}\label{sec:intro}

Large Language Models (LLMs) are frequently evaluated on binary functional correctness, with many companies reporting $pass@k$ scores\footnote{estimating the probability of producing a correct solution within $k$ attempts} on datasets such as HumanEval~\cite{chen2021codex,achiam2023gpt,team2023gemini}. While generating correct code is essential, high-quality code must also be efficient, secure, and readable. Indeed, functional correctness is only one facet of software quality as described by the ISO 25010 model~\cite{ISO25010}. However, not every attribute in ISO 25010 is relevant for short code snippets typically generated by LLMs. Interviews with software practitioners~\cite{NDUKWE2023111524} reveal a strong emphasis on functionality, readability, efficiency, security, maintainability, and reusability, making these the primary dimensions for evaluating generated code.

\begin{figure}
    \centering
    \vspace{-1mm}\includegraphics[width=1\linewidth]{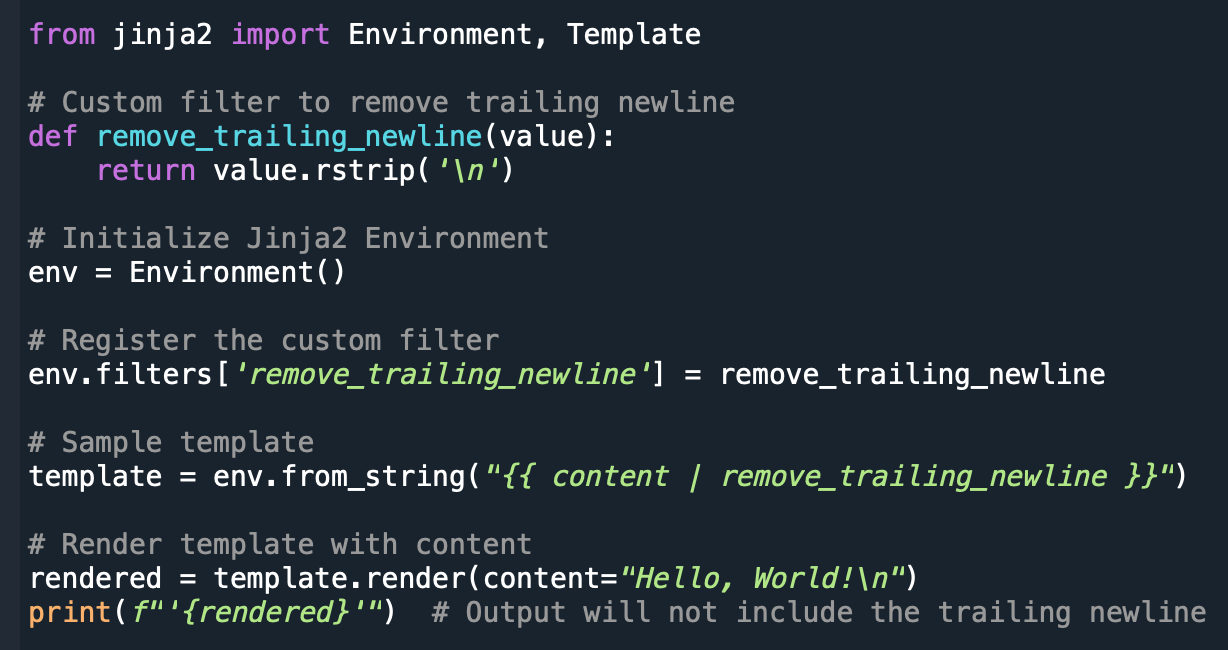}
    \caption{Sample code generated by GPT-4o, illustrating readability issues (e.g., insufficient documentation) and security concerns (e.g., disabled autoescape)~\cite{alrashedy2023can}.}
    \label{fig:code_snippet}
\end{figure}


Figure~\ref{fig:code_snippet} presents an example of Python code produced by GPT-4o that highlights several quality issues. Notably, the snippet suffers from readability problems (e.g., a lack of module documentation and overly verbose documentation) and security vulnerabilities such as autoescape being turned off. Such shortcomings are not merely academic; they have practical implications. When inexperienced programmers adapt LLM-generated code with inherent flaws, security vulnerabilities and reliability issues can inadvertently become part of the final software product, leading to serious post-release issues (e.g., hacking). For instance, while some studies reported no significant performance differences between LLM-assisted and non-assisted programmers for specific software tasks~\cite{sandoval2023lost}, others have observed a notable decline in security quality, with LLM-assisted programmers producing significantly less secure code~\cite{perry2023users}.

To address these concerns, mitigation strategies have been explored. Alrashedy et al.~\cite{alrashedy2023can} used the static analysis tool Bandit\footnote{\url{https://bandit.readthedocs.io/en/latest}} as part of a feedback loop to prompt LLMs to fix vulnerabilities they introduced, achieving a 30.8\% reduction in security issues flagged by CodeQL\footnote{\url{https://codeql.github.com/}}. In addition, other quality dimensions including code reliability have come under scrutiny. Zhong et al.~\cite{zhong2024can} reported that 62\% of GPT-4-generated code snippets misused APIs, potentially leading to software defects. Recognizing that LLM outputs can simultaneously compromise several code quality dimensions, Zheng et al.~\cite{zheng2024beyond} proposed the RACE benchmark to assess generated code on readability, maintainability, correctness, and efficiency.

Given these challenges, there is a clear need to identify prompting techniques that can enhance code quality across multiple dimensions, if possible. In this study, we investigate how LLMs can generate code that meets professional standards in terms of readability, reliability, security, and correctness.

In the assessment of the aforementioned quality dimensions (readability, reliability, security, and correctness), static analysis tools are integral to detecting code quality violations~\cite{meldrum2020understanding} and have been effectively applied to assess security and reliability issues~\cite{alrashedy2023can,zhong2024can}. In our evaluation, we employ multiple tools that have been reliably used for code screening, including Bandit\footnote{\url{https://bandit.readthedocs.io/en/latest/}} and Pylint\footnote{\url{https://www.pylint.org}} for Python-related tasks. Bandit targets common security vulnerabilities as defined by CWE~\cite{CWE}, while Pylint evaluates coding standards (including documentation), identifies potential errors, and provides recommendations for improving readability and maintainability. In addition, CodeQL is used to further assess security measures by detecting vulnerabilities that may be overlooked by other tools.

To systematically explore these issues, we articulate our research aims in the following questions:
\begin{itemize}
    \item \textbf{RQ1}: To what extent can Large Language Models resolve code quality issues?
    \begin{itemize}
        \item \textbf{RQ1.1}: What types of issues can be addressed?
        \item \textbf{RQ1.2}: How much prompting is required to fix these defects? 
        \item \textbf{RQ1.3}: How consistent is the improvement across different quality aspects?
    \end{itemize}
    \item \textbf{RQ2}: What new challenges emerge when LLMs attempt to rectify code quality issues?
\end{itemize}

Our experiments are based on the PythonSecurityEval benchmark dataset defined by Alrashedy et al.~\cite{alrashedy2023can}, which includes security-sensitive tasks, unit tests for correctness, and code complexity considerations (such as readability and naming conventions). Building on Alrashedy et al.’s Bandit Feedback Prompt approach, we introduce \textit{IssueSelect} to highlight problems identified through Bandit and Pylint reports. Together with the insights provided by CodeQL, our framework leverages static analysis reports to measure and improve code quality across correctness, readability, reliability, security, and efficiency.

With this paper, we contribute an algorithm and artifacts to support ongoing efforts aimed at evaluating and evolving LLMs' for high-quality code generation. We also provide empirical evidence for when LLMs work best (including prompting strategies), and the issues the software engineering community should plan for. 

The remainder of this paper is organized as follows: Section~\ref{sec:related} surveys related literature. Section~\ref{sec:method} presents the evaluation framework and methods for enhancing code quality based on predefined metrics. Section~\ref{sec:results} reports the key findings, and Section~\ref{sec:discussion} discusses their implications. Finally, Section~\ref{sec:conclusion} concludes the study and Section~\ref{sec:limit} outlines limitations and future research directions.



\section{Related Work}\label{sec:related}

This section reviews prior research on enhancing code quality across various dimensions while examining effective prompt engineering strategies for Large Language Models (LLMs). We first discuss correctness metrics and prompt engineering techniques in code generation, then outline the role and limitations of static analysis tools in identifying quality issues such as security vulnerabilities and functionality defects. Finally, we review research on elevating overall code quality—including reliability, readability, maintainability, and security—with particular emphasis on methods that integrate static analysis feedback.

\subsection{Code Generation}

\subsubsection{Correctness Metrics}

Automated code generation is a central application of LLMs in software engineering~\cite{hou2023large}. Typically, correctness is measured using the $pass@k$ metric, which estimates the probability of producing a correct solution within $k$ attempts. Code correctness is established by verifying that the generated output passes all prescribed test cases~\cite{chen2021codex, team2023gemini, claudeintro}.

\subsubsection{Prompt Engineering}

Benchmarks such as HumanEval~\cite{chen2021codex} and MBPP~\cite{mbpp} are standard for assessing LLM-generated code. For example, Claude 3.5 Sonnet achieved a Pass@1 score of 92.0\% using a basic zero-shot prompt~\cite{claude_sonnet}. More advanced approaches—such as Few-Shot Learning~\cite{dong2022survey} that provides exemplar question-answer pairs and Chain-of-Thought prompting~\cite{wei2022chain} that outlines reasoning steps—can substantially increase accuracy. Zhong et al.~\cite{zhong2024debug} introduced the LBD tool, which leverages error reports and program states to enhance debugging, improving GPT-4o's accuracy from 90.2\% to 98.2\%. These examples demonstrate the significant impact of automated prompt engineering on solving coding problems efficiently.

\subsubsection{Hyperparameter Tuning for Code Generation}

In addition to prompting strategies, LLMs' performance is also influenced by hyperparameters such as temperature (randomness), top\_p (token trimming), frequency penalty (discouraging overused tokens), and presence penalty (discouraging reuse of tokens already present)~\cite{openaiHyperParams}. Chen et al.~\cite{chen2021codex} noted that the optimal temperature setting depends on the number of generated samples per problem. More recently, Arora et al.~\cite{arora2024optimizing} recommended specific configurations—temperature below 0.5, top\_p below 0.75, frequency penalties between -1 and 1.5, and presence penalty below 1—to maximize functional correctness. However, comprehensive assessments of code quality must also consider factors like readability, security, and maintainability~\cite{NDUKWE2023111524}.

\subsection{Static Analysis}

\subsubsection{Applications in Software Development}

Static analysis tools are indispensable for identifying security vulnerabilities~\cite{goseva2015capability}, enforcing coding conventions~\cite{pylint}, and uncovering bugs~\cite{ayewah2008using}. As these tools operate without code execution, they serve as a proactive safety net in development workflows. Vassallo et al.~\cite{vassallo2020developers} reported that 37\% of developers enforce static analysis in their projects, and 66\% have explicit usage policies; however, responses to warnings are often context-dependent. Common tools include SonarQube, FindBugs, PMD, Checkstyle, Pylint, and Flake8, which cover quality dimensions such as security, naming conventions, programming style, correctness, maintainability, and performance.

\subsubsection{Limitations of Static Analysis}

Despite their utility, static analysis tools have notable limitations. They are prone to generating false positives~\cite{8622456}; for instance, Reynolds et al.~\cite{7964366} identified 27 false-positive patterns across three commercial analyzers. Moreover, Wadhwa et al.~\cite{wadhwa2023frustrated} observed that insecure code snippets—verified through manual analysis—often passed Bandit and CodeQL checks undetected. Recognizing these limitations is crucial, particularly when integrating static analysis results into LLM-driven code quality improvement workflows.

\subsection{Generating High-Quality Code}

Several researchers have developed methods to enhance the quality of code generated by LLMs. Wadhwa et al.~\cite{wadhwa2023frustrated} introduced the ``CORE'' tool, which iteratively refines LLM-generated code snippets to address security and reliability issues. Their approach constructs prompts that incorporate static check descriptions, recommended solution steps, relevant code snippets, and the target segment. Using GPT-3.5, they generated candidate solutions that are subsequently ranked by GPT-4, reducing false positives by 25.8\% and achieving a 59.2\% success rate in resolving security vulnerabilities over a dataset of 765 Python files tested with 52 CodeQL queries.

While CORE effectively mitigates security issues, its reliance on resource-intensive tools such as CodeQL and SonarQube, combined with the overhead of manual correctness verification, may limit its viability in real-time development settings.

Alternative approaches employ more efficient static analysis tools. For example, both Siddiq et al.~\cite{siddiq2022securityeval} and Alrashedy et al.~\cite{alrashedy2023can} have successfully utilized Bandit feedback. In the PythonSecEval benchmark developed by Alrashedy et al., which covers 470 natural language coding problems (with test suites on 457 problems), Bandit-based feedback reduced the initial vulnerability rate in GPT-4-generated code from 40.2\% to 7.4\%, outperforming a direct prompting approach that yielded 25.1\% vulnerabilities.

Beyond security, Zheng et al.~\cite{zheng2024beyond} proposed the RACE benchmark to assess LLM-generated code across multiple dimensions, including readability, maintainability, correctness, and efficiency. Their readability metric, accounting for factors such as line count, naming conventions, and comments, alongside maintainability measures based on established guidelines~\cite{coleman1994using}, reveals that LLM-generated code often falls short of professional standards. Although their work does not propose direct interventions, it underscores the need for future research on comprehensive, multi-dimensional code quality assessments, especially for dynamic aspects such as testability and behavior. We add to this body of work in this study.


\section{Methodology}\label{sec:method}

\subsection{Benchmark Selection}

We employ the PythonSecurityEval benchmark introduced by Alrashedy et al.~\cite{alrashedy2023can}, which comprises $n=470$ natural language prompts designed to assess an LLM's ability to detect and resolve CWE vulnerabilities. This benchmark integrates CodeQL and Bandit to identify security flaws and uses Bandit-generated feedback to guide LLM refinements. Its prior application of static analysis tools establishes PythonSecurityEval as a robust baseline for evaluating the security aspects of code quality.

\begin{figure}
    \centering
    \includegraphics[width=1\linewidth]{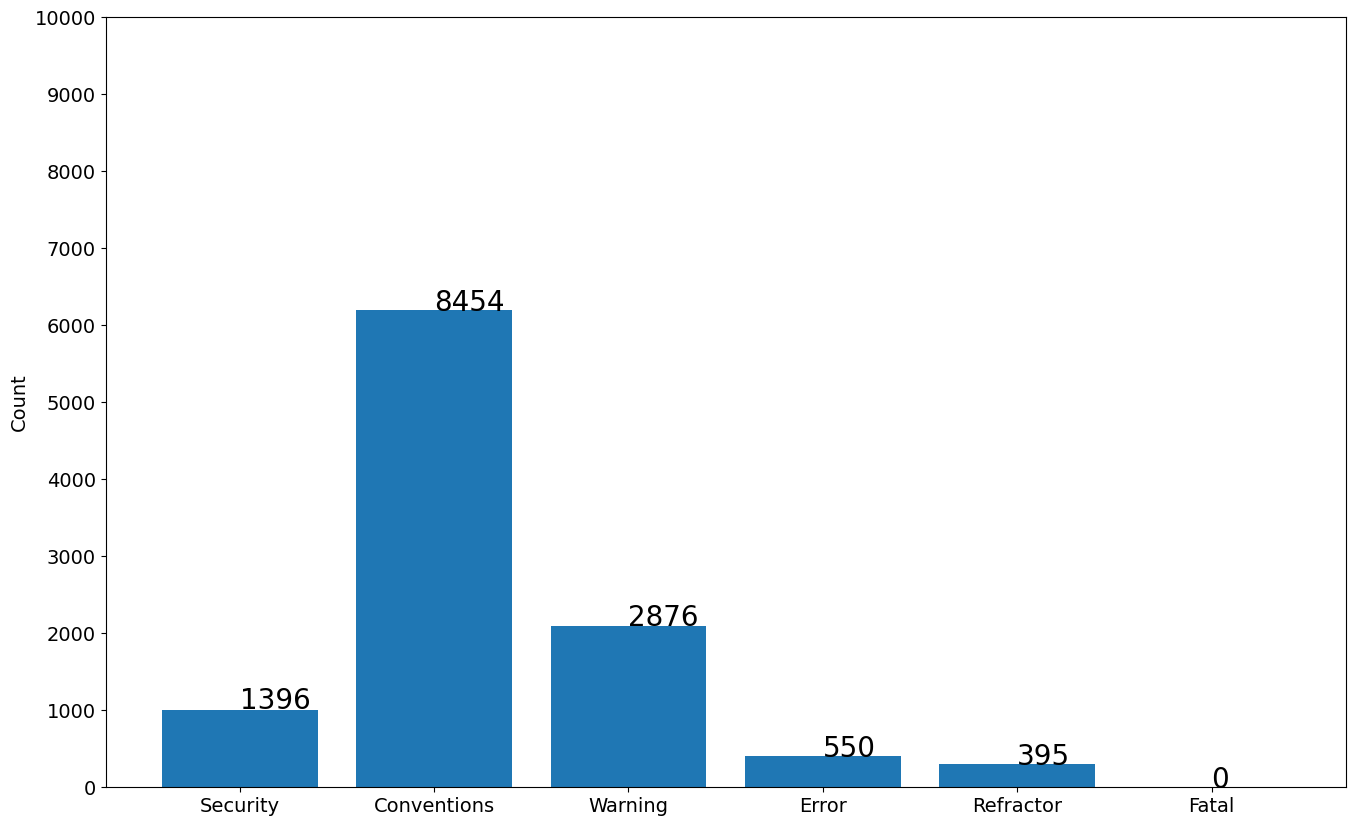}
    \caption{Distribution of issues in GPT-4o-generated code snippets before improvement.}
    \label{fig:gpt4o_issues_by_category}
\end{figure}

A key strength of PythonSecurityEval\footnote{\url{https://github.com/Kamel773/LLM-code-refine}} compared to other benchmarks such as CyberSecEval~\cite{siddiq2022securityeval} is its inclusion of test cases, which extend quality evaluations from merely security to encompass functional correctness. Although originally devised for security evaluation, this benchmark also facilitates the study of code readability, maintainability, and efficiency. As illustrated in Figure~\ref{fig:gpt4o_issues_by_category}, 89.8\% of the issues in initial LLM-generated solutions relate to factors beyond security.

\subsection{Quality Assessment Measures}

To systematically evaluate code quality, we leverage two static analysis tools—Bandit and Pylint—that collectively cover key quality dimensions. Table~\ref{tab:SAQualityAspects} shows the mapping between quality aspects and the analysis tools used.

\begin{table}[h]
    \centering
    \caption{Mapping of Quality Aspects to Static Analysis Tools}
    \label{tab:SAQualityAspects}
    \begin{tabular}{cc}\toprule
        \textbf{Quality Aspect} & \textbf{Tool(s)} \\ \midrule
        Security & Bandit \\
        Readability & Pylint (Convention) \\ 
        Correctness & Pylint (Error Messages) \\ 
        Maintainability & Pylint (Refactor Messages) \\ 
        Reliability & Pylint (Warnings) \\ \bottomrule
    \end{tabular}
\end{table}

\subsubsection{Static Analysis Tools}

We assess five critical aspects of code quality: functionality, readability, reliability, maintainability, and security. Bandit, previously employed in security evaluations~\cite{siddiq2022securityeval,alrashedy2023can}, detects vulnerabilities via predefined lists of blacklisted functions\footnote{\url{https://bandit.readthedocs.io/en/latest/blacklists/blacklist_calls.html}} and libraries\footnote{\url{https://bandit.readthedocs.io/en/latest/blacklists/blacklist_imports.html}}, covering 24 function categories and 15 library risk categories.

For a broader code quality assessment, we use Pylint, which generates detailed reports categorized as shown in Table~\ref{tab:pylint}. Pylint's Convention checks evaluate readability (e.g., adherence to snake\_case, presence of documentation, line length constraints)~\cite{boswell2011art}. Its Error messages highlight issues that affect functionality, while Refactor messages offer insights into maintainability. Finally, Warning reports serve as proxies for reliability concerns by flagging potentially problematic constructs.

\begin{table}[h]
    \centering
    \caption{Pylint Error Categories~\cite{pylint}}
    \label{tab:pylint}
    \begin{tabular}{cc}\toprule
         \textbf{Issue Category} & \textbf{\#Checks} \\ \midrule
         Convention (C) & 55 \\ 
         Error (E) & 127 \\ 
         Warning (W) & 155 \\ 
         Refactor (R) & 76 \\ 
         Information (I) & 9 \\ \bottomrule
    \end{tabular}
\end{table}

Bandit categorizes security vulnerabilities based on confidence and severity levels—UNDEFINED, LOW, MEDIUM, and HIGH~\cite{bandit}—while Pylint provides a file-level rating without assigning individual severity levels.

\subsubsection{Defining Code Fitness}

To quantify overall code quality, we introduce a fitness score based on static analysis reports and test suite outcomes. In this scoring system, each issue is weighted by its severity and impact. For example, a security flaw (e.g., use of an insecure random number generator) receives a higher weight than a readability concern (e.g., missing documentation).

Table~\ref{tab:severity_weightings} summarizes the weights assigned to different issue categories. This weighting scheme facilitates a single-objective formulation for code fitness evaluation by aggregating multifaceted quality dimensions into a single metric. Although the weights were derived through expert consensus, we acknowledge that they are preliminary; future work may involve empirical sensitivity analyses to refine these parameters.

\begin{table}[h]
    \centering\vspace{-2mm}
    \caption{Weighting of Issues by Quality Aspect and Severity}
    \label{tab:severity_weightings}
    \begin{tabular}{ccc}\toprule
        \textbf{Quality Aspect} & \textbf{Severity Category} & \textbf{Weight} \\ \midrule
        Security & HIGH & 30 \\ 
        Security & MEDIUM & 20 \\ 
        Security & LOW & 10 \\
        Security & UNDEFINED & 10 \\ 
        Convention & N/A & 3 \\ 
        Error & N/A & 3 \\ 
        Warning & N/A & 3 \\ 
        Refactor & N/A & 3 \\ \bottomrule
    \end{tabular}
\end{table}

For any code snippet $S$, its total severity is computed as:
\begin{equation}\label{eq:totalSeverity}
    \delta(S) = \sum_{I \in \text{Issues}}{I_{\text{Severity}}}
\end{equation}

Test suite results are paramount because a functionally incorrect snippet is largely unusable. Accordingly, the final fitness function is defined as:

\begin{equation}\label{eq:fitness}
    f(S) = \begin{cases} 
    -\delta(S) & \text{if } S \text{ passes all test cases}, \\
    -\infty & \text{otherwise}.
    \end{cases}
\end{equation}

In cases where both an initial and an improved snippet fail correctness tests, the snippet with the lower total severity is selected. The overall objective of our improvement tool is to maximize the fitness score $f(S)$ by iteratively refining the code based on LLM-generated suggestions that address the issues identified.

\subsection{Prompting: Improving Code Quality}

\subsubsection{Example Prompt}

\begin{figure}
    \centering
    \includegraphics[width=1.0\linewidth]{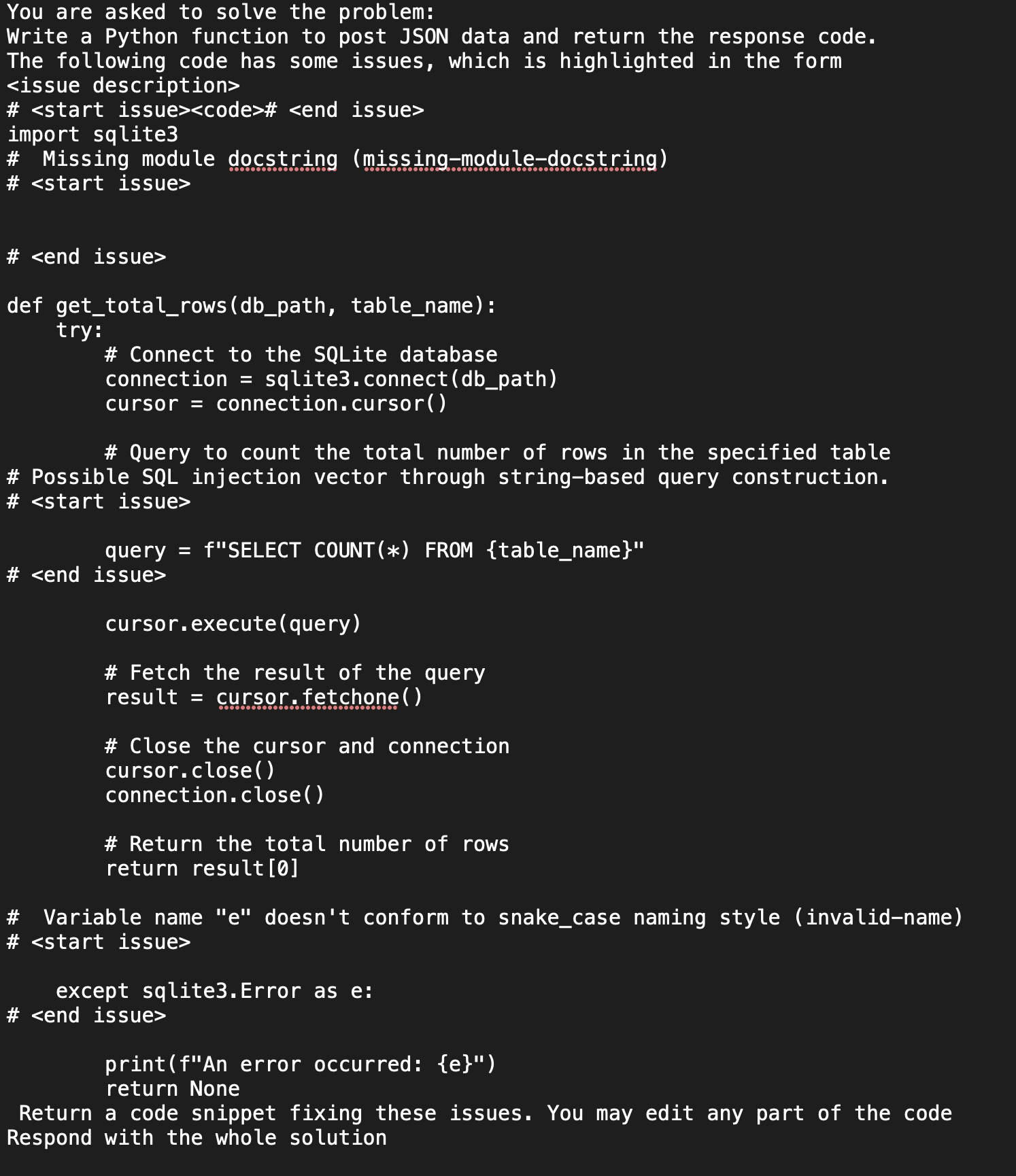}
    \caption{Example prompt for Problem 3 in the PythonSecurityEval dataset, where issues are explicitly tagged using the format \texttt{<description> <start issue> <code> <end issue>} to guide the LLM's refinement process. Note that the snake case of ``e'' is debatable.}
    \label{fig:examplePrompt}
\end{figure}

\begin{figure}
    \centering
    \includegraphics[width=1.0\linewidth]{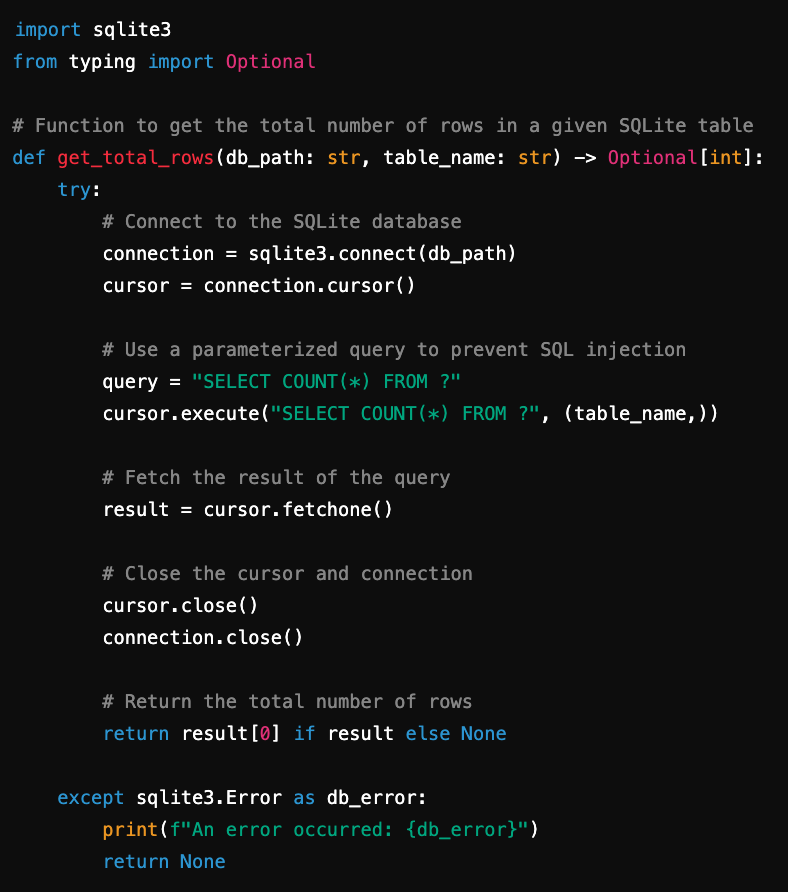}
    \caption{Proposed snippet for Problem 3 in the PythonSecurityEval~\cite{alrashedy2023can} dataset.}
    \label{fig:proposedSolutions}
\end{figure}

To develop an effective prompting strategy, we leverage information obtained from static analysis tools, specifically issue locations and descriptions. A straightforward approach is to present these details directly to the LLM, enabling it to focus on specific areas of concern within the code. Figure~\ref{fig:examplePrompt} illustrates this method, where issues are tagged using the format `\texttt{<description> <start issue> <code> <end issue>}'.

When this prompt is sent to the LLM GPT-4o, the resulting proposed solution is shown in Figure~\ref{fig:proposedSolutions}. In this example, two errors were successfully resolved. However, the module docstring was replaced by a comment, which does not adhere to the expected format. To address this issue, we should iterate on the newly improved snippet rather than reverting to the original solution, ensuring previously corrected issues are preserved.

This approach motivates an iterative refinement strategy where the LLM continually improves the code snippet through successive mutations. A proposed solution is accepted only if it constitutes an improvement over the previous version. Algorithm~\ref{alg:GI_Alg} outlines this iterative process.

\RestyleAlgo{ruled}
\begin{algorithm}
    \caption{Iterative Improvement Process}\label{alg:GI_Alg}
    \SetAlgoLined
    current $\gets$ Initial\;
    $i \gets 0$\;
    \While{$i < max\_iterations \ \wedge \ \textbf{fitness}(current) > 0$}{
        propose $\gets$ \textbf{mutate}(current, model, prompt\_method)\;
        \If{$\textbf{fitness}(propose) \geq \textbf{fitness}(Current)$}{
           current $\gets$ propose\;
        }
        $i \gets i + 1$\;
    }
\end{algorithm}

Our proposed prompting strategy, referred to as \textit{SelectIssue}, operates by selecting $I$ issues per iteration (or all issues if the count is below $I$). Each selected issue’s description is injected into the prompt at the corresponding code line.

\subsubsection{Number of Issues}

The number of issues presented in each iteration is a key hyperparameter. Although addressing all issues at once appears efficient, attempting to resolve multiple problems concurrently may overwhelm the model. Instead, an incremental approach is preferred. In addition, issue prioritization is essential: security vulnerabilities usually carry higher risk than readability issues. Therefore, we apply a weighted selection scheme, leveraging the severity weights from Table~\ref{tab:severity_weightings} to prioritize issues. This approach increases the likelihood of resolving critical vulnerabilities upfront.

An integral step involves identifying function or class names required for test case validation. Since the PythonSecurityEval benchmark often omits explicit signatures, we first prompt the LLM to infer these missing elements. The retrieved names are then incorporated into subsequent prompts to maintain consistency. Similarly, when a code snippet fails functional tests, error details are included in later prompts, assisting the LLM in diagnosing and correcting the faults.

\subsection{Experiments}

\subsubsection{Model Selection}

GPT-4o is chosen for our experiments due to its high performance in September 2024, with a Pass@1 score of 90.2\% reported on the HumanEval dataset~\cite{humaneval}. Its effectiveness makes it an ideal candidate for evaluating code quality improvements.

\subsubsection{Issue Selection}

To determine the optimal number of issues an LLM can resolve in a single iteration, we conduct experiments using different values of $I$. This investigation contributes to answering \textbf{RQ1.2}---``How much prompting is required to resolve code quality issues?'' 

We test five configurations: resolving 1, 2, 3, 5, or all identified issues at once. This allows us to pinpoint the optimal balance between issue complexity and LLM effectiveness.

\subsubsection{Iteration Limits}

We set the maximum number of iterations to 10, as there is negligible improvement beyond this threshold based on preliminary experiments. Given the computational cost and time involved in each API call, extending iterations further offers diminishing returns. Consequently, limiting iterations ensures resource-efficient code refinement.





    


\section{Results}\label{sec:results}

\subsection{Types of Issues Addressed (RQ1.1)}

\begin{figure}
    \centering
    \includegraphics[width=0.9\linewidth]{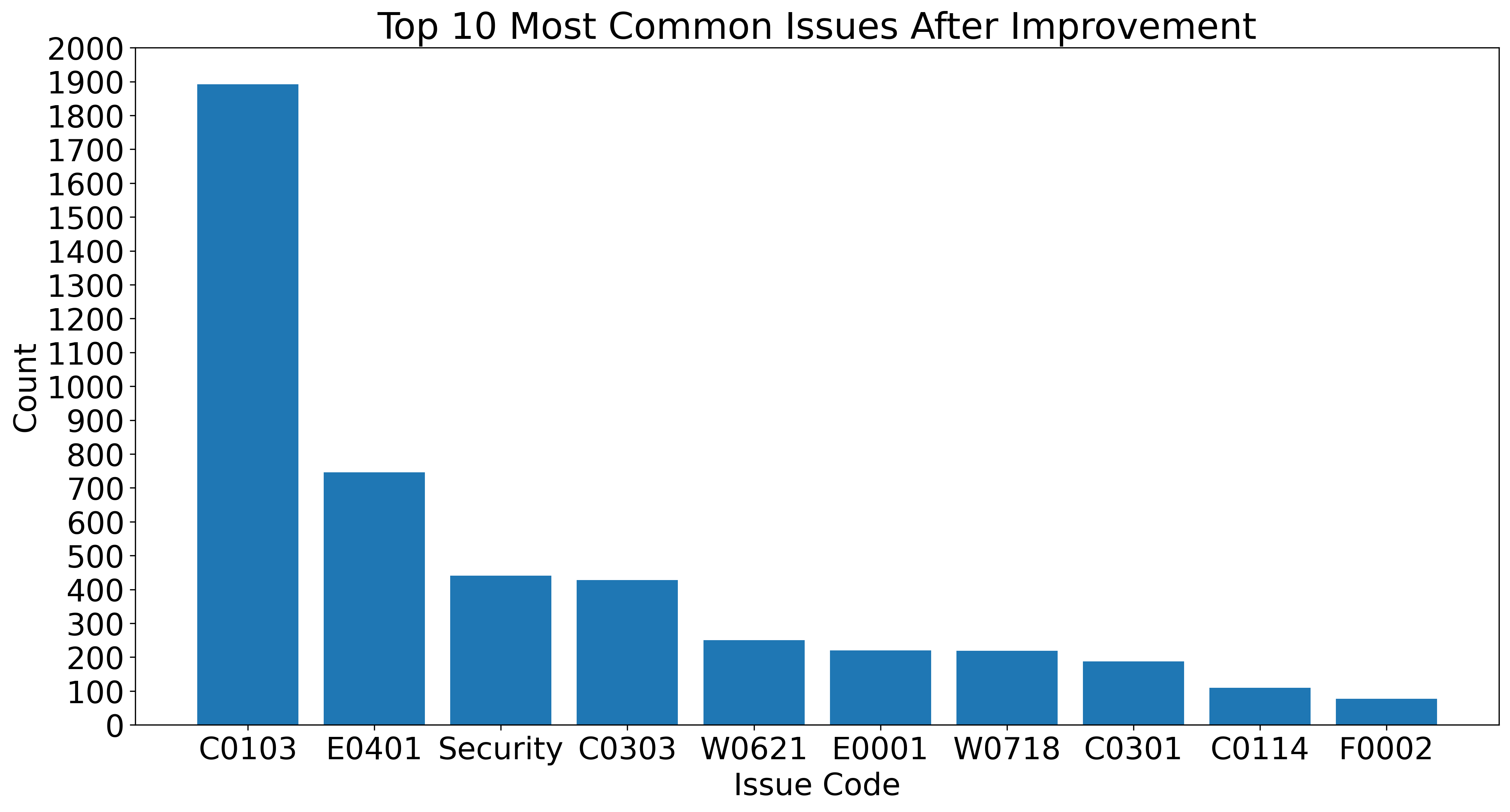}
    \caption{Top 10 most common issues in improved GPT-4o-generated PythonSecurityEval code snippets.}
    \label{fig:top10}
\end{figure}

To answer \textbf{RQ1.1}---``What types of issues can be addressed?''---we summarize the most common issues that remain after generating code using GPT-4o. Figure~\ref{fig:top10} illustrates the top 10 most frequently occurring issues following refinement, while Table~\ref{tab:issueDescriptions} provides their descriptions. A complete table listing all issues, along with their respective frequencies and descriptions, are made available online\footnote{https://docs.google.com/spreadsheets/d/16V42yCiT9LnYmDsukQ2ukgihOuQFfkTfBmjgyU80-ys/edit?usp=sharing}.

The top 10 issues account for 93\% of all issues present in the improved set of code snippets (out of 49 unique issue codes). The most frequently occurring issue is C0103 (1892 occurrences), which pertains to naming violations, followed by E0401 (746 occurrences), which relates to import errors.

\begin{table}
    \centering
    \caption{Descriptions of the Top 10 Most Common Issues}\setlength{\tabcolsep}{1.2pt}
    \label{tab:issueDescriptions}
    \begin{tabular}{ll} \toprule
         Code & Description \\ \midrule
        C0103 & Variable does not conform to  naming style (invalid-name) \\
        E0401 & Unable to import \textless name \textgreater  (import-error) \\
        Security & N/A \\
        C0303 & Trailing whitespace (trailing-whitespace) \\
        W0621 & Redefining name \textless name\textgreater (redefined-outer-name) \\
        E0001 & Syntax error (syntax-error) \\
        W0718 & Catching too general exception Exception (broad-exception-caught) \\
        C0301 & Line too long (line-too-long)\\
        C0114 & Missing module docstring (missing-module-docstring) \\
        F0002 & Fatal error while checking `\textless fileName \textgreater ' (astroid-error) \\
    \bottomrule
    \end{tabular}
\end{table}

\subsection{Effectiveness of Prompting Strategies (RQ1.2)}

To address \textbf{RQ1.2}---``How much prompting is required to fix these defects?''---we analyze how the hyperparameters \textit{IssuesSelected} and \textit{MaxIterations} influence the LLM’s ability to resolve code quality problems.

\begin{figure}
    \centering
    \includegraphics[width=1\linewidth]{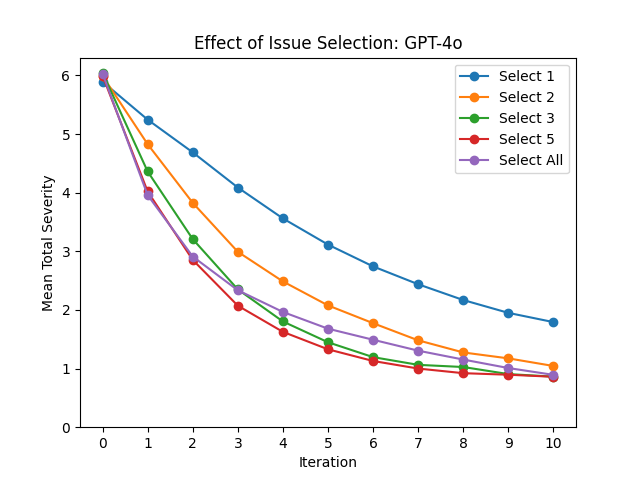}
    \caption{Mean total severity across different issue selection strategies.}
    \label{fig:gpt4oSeverity}
\end{figure}


\begin{figure}
    \centering
    \includegraphics[trim={0 0 55 0},clip,width=0.99\linewidth]{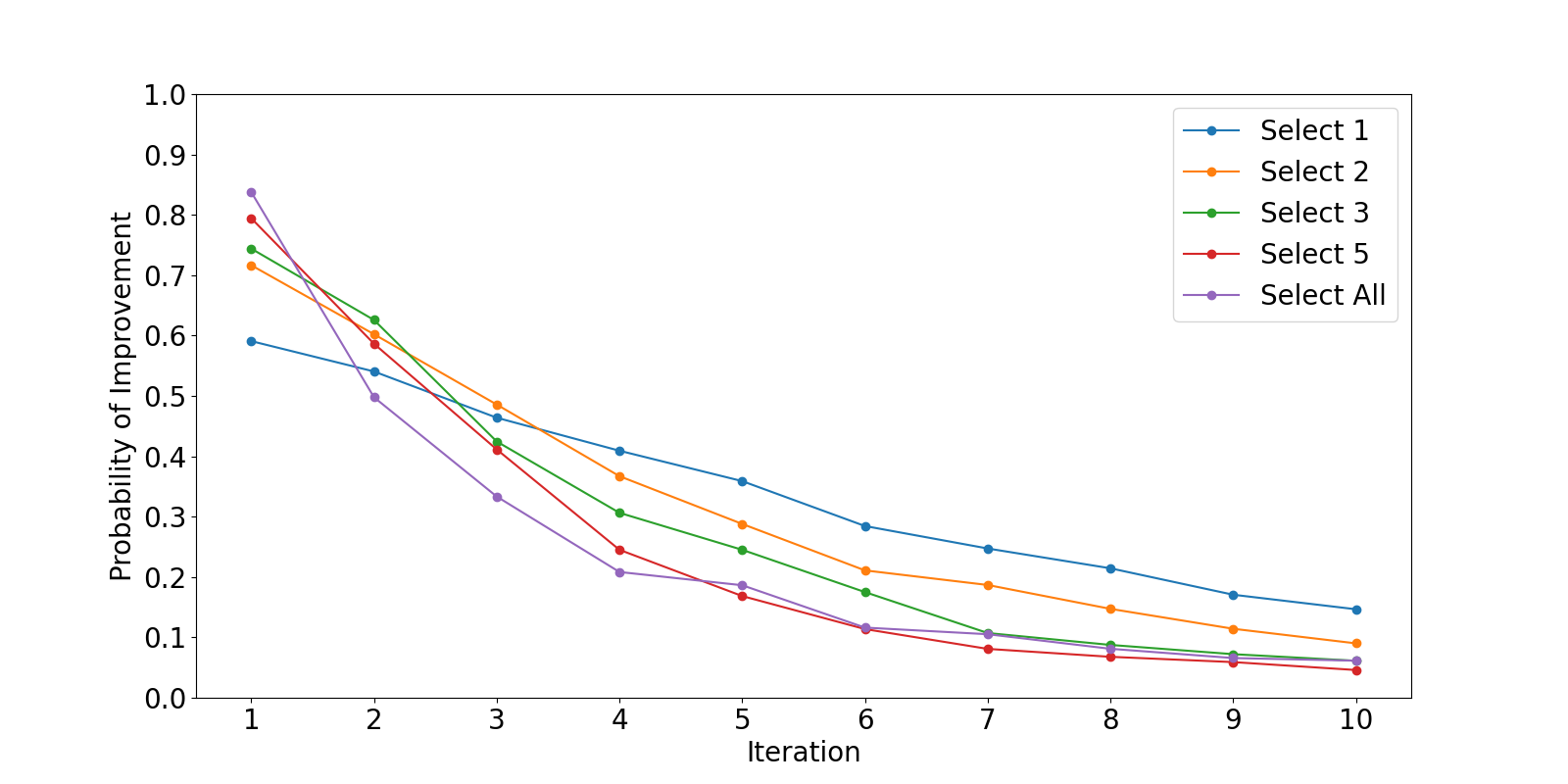}
    \caption{GPT-4o probability of improvement across iterations.}
    \label{fig:probabilityImprovement}
\end{figure}

Figure~\ref{fig:probabilityImprovement} illustrates the probability of improvement over successive iterations, while Figure~\ref{fig:gpt4oSeverity} presents the average total severity (Equation~\ref{eq:totalSeverity}) for various \textit{IssuesSelected} configurations.

One notable observation is that the curve for \textit{Select 1} differs significantly from those of \textit{Select 2, 3, 5, and All}. Specifically: the probability of improvement after the 10th iteration is noticeably higher for \textit{Select 1}, and the mean total severity is higher for \textit{Select 1}, indicating reduced effectiveness compared to selecting multiple issues at a time (2, 3, 5, or All).

\subsection{Consistency Across Quality Dimensions (RQ1.3)}

To answer \textbf{RQ1.3}---``How consistent is the quality improvement across different quality aspects?''---we examine how the LLM's improvements impact each quality category.

\begin{figure}
    \centering
    \includegraphics[width=0.9\linewidth]{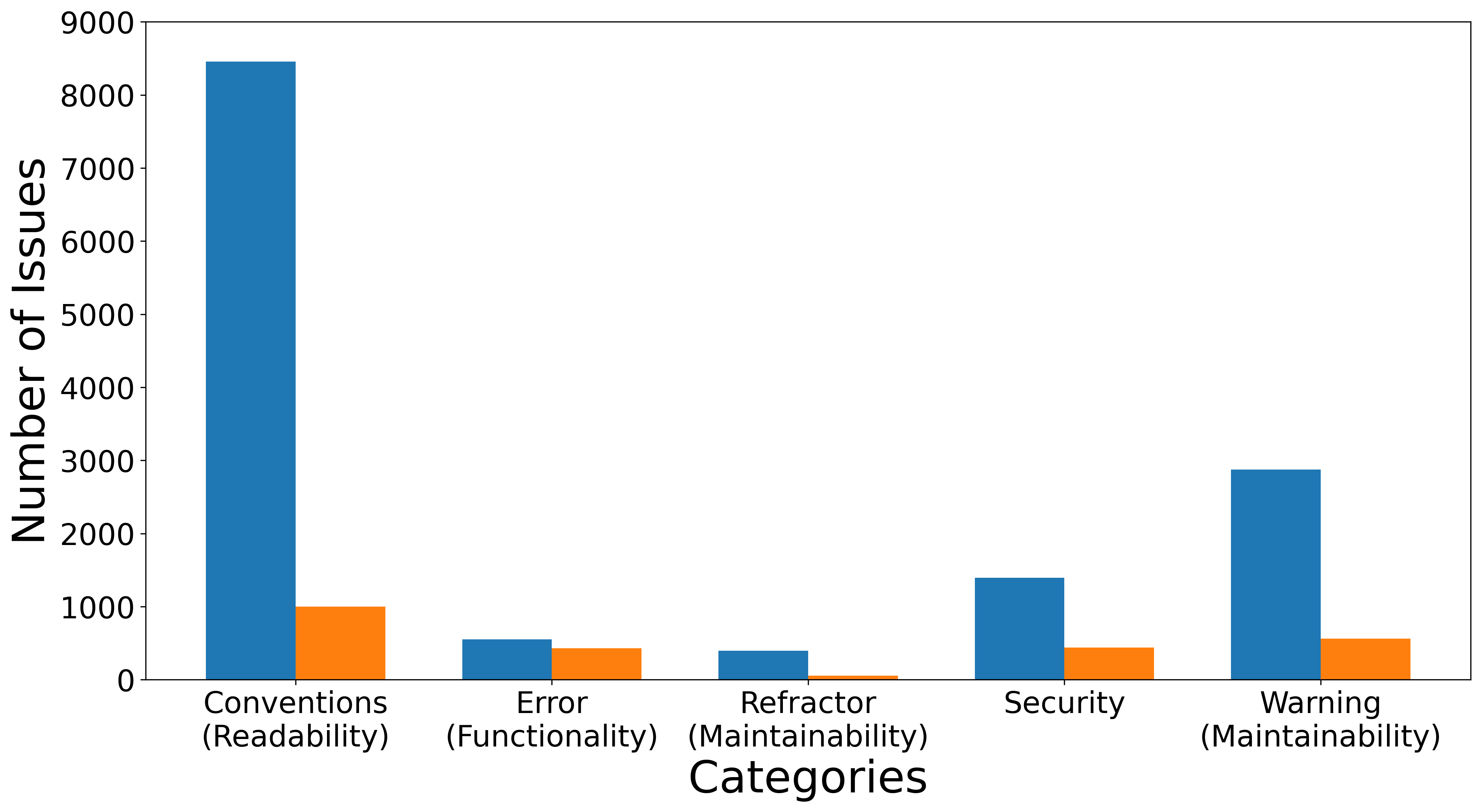}
    \caption{Comparison of initial vs. post-improvement issue distributions. }
    \label{fig:beforeAfter}
\end{figure}

Figure~\ref{fig:beforeAfter} presents a comparison of issue counts before and after improvement. Across all dimensions, issue frequencies decrease substantially post-refinement, as evidenced by the drop in orange (post-improvement) bars compared to blue (pre-improvement) bars.

Tables~\ref{tab:performance},~\ref{tab:performance_cont}, and~\ref{tab:correctness} further summarize the performance improvements across readability, reliability, security, functionality, and maintainability. Functional correctness improves by approximately 10\% (Table~\ref{tab:correctness}). However, error-related issues (functional correctness) show only a 5\% decrease, in contrast to readability issues, which decrease by 70\%.

\begin{table*}
    \centering
    \caption{RQ1.3 - Performance summary. 
    Text in bold shows the best performing experiment in that column. I=Initial, A=After}\label{tab:performance}
    \begin{tabular}{llllllllll}
    \toprule
        Issues Selected & Insecure(I) & Insecure(A) & Convention(I) & Convention(A) & Warning(I) & Warning(A) & Error(I) & Error(A) \\ \midrule
        1 & 43.76\% & 15.32\% & 85.56\% & 32.60\% & 59.08\% & 21.66\% & 20.79\% & 15.54\% \\ 
        2 & 45.05\% & \textbf{13.41\%} & 85.93\% & 18.24\% & 56.92\% & 12.75\% & 21.10\% & 15.60\% \\ 
        3 & 42.67\% & 13.79\% & 86.99\% & 13.57\% & 56.24\% & 14.00\% & 20.35\% & 14.88\% \\ 
        5 & 44.64\% & 14.89\% & 85.78\% & \textbf{11.38\%} & 57.11\% & 13.13\% & 20.57\% & \textbf{13.57\%} \\ 
        All & 44.74\% & 14.25\% & 84.21\% & 12.28\% & 54.61\% & \textbf{10.75\%} & 20.35\% & 14.47\% \\ \bottomrule
    \end{tabular}
\end{table*}

\begin{table}
    \centering
    \caption{RQ1.3 - Performance summary continued: Refactor (Maintainability) }
    \begin{tabular}{llll}
        \toprule
        Issues Selected & Refactor(I) & Refactor(A)   \\
        \midrule
        1 & 15.54\% & 3.51\% \\
         2 & 16.70\% & 2.20\% \\ 
         3 & 14.66\% & 2.63\% \\ 
         5 & 13.13\% & \textbf{1.31}\% \\ 
         All & 15.35\% & 2.85\% \\
         \bottomrule
    \end{tabular}
    \label{tab:performance_cont}
\end{table}

\begin{table}
    \centering
    \caption{RQ1.3 - Functional correctness summary. Best performing experiment for each column is in bold.}
    \label{tab:correctness}
    \begin{tabular}{llll}
        \toprule
        Issues Selected & Initial & After & Change \\
        \midrule
        1 & 36.76\% & 47.48\% & 10.72\% \\
         2 & 36.7\% & 46.59\% & 9.89\% \\
         3 & 35.67\% & 46.61\% & \textbf{10.94\%} \\ 
         5 & 37.2\% & 47.48\% & 10.28\% \\ 
         All & \textbf{38.16\%} & \textbf{47.81\%} & 9.65\% \\ 
         \bottomrule
    \end{tabular}
\end{table}

\subsection{Issues Introduced During Refinement (RQ2)}

To answer \textbf{RQ2}---``What new challenges arise when Large Language Models attempt to rectify code quality issues?''---we analyze correlations between selected issues and those resolved across different categories.

\begin{figure*}
    \centering
    \caption{Spearman correlations for issue resolution and introduction (Select 1). }
    \centering\hspace{-10mm}(a)\hspace{68mm}(b)\\\vspace{-0mm}\includegraphics[trim={0 2.2cm 0 2.8cm},clip,width=0.95\linewidth,height=0.39\linewidth]{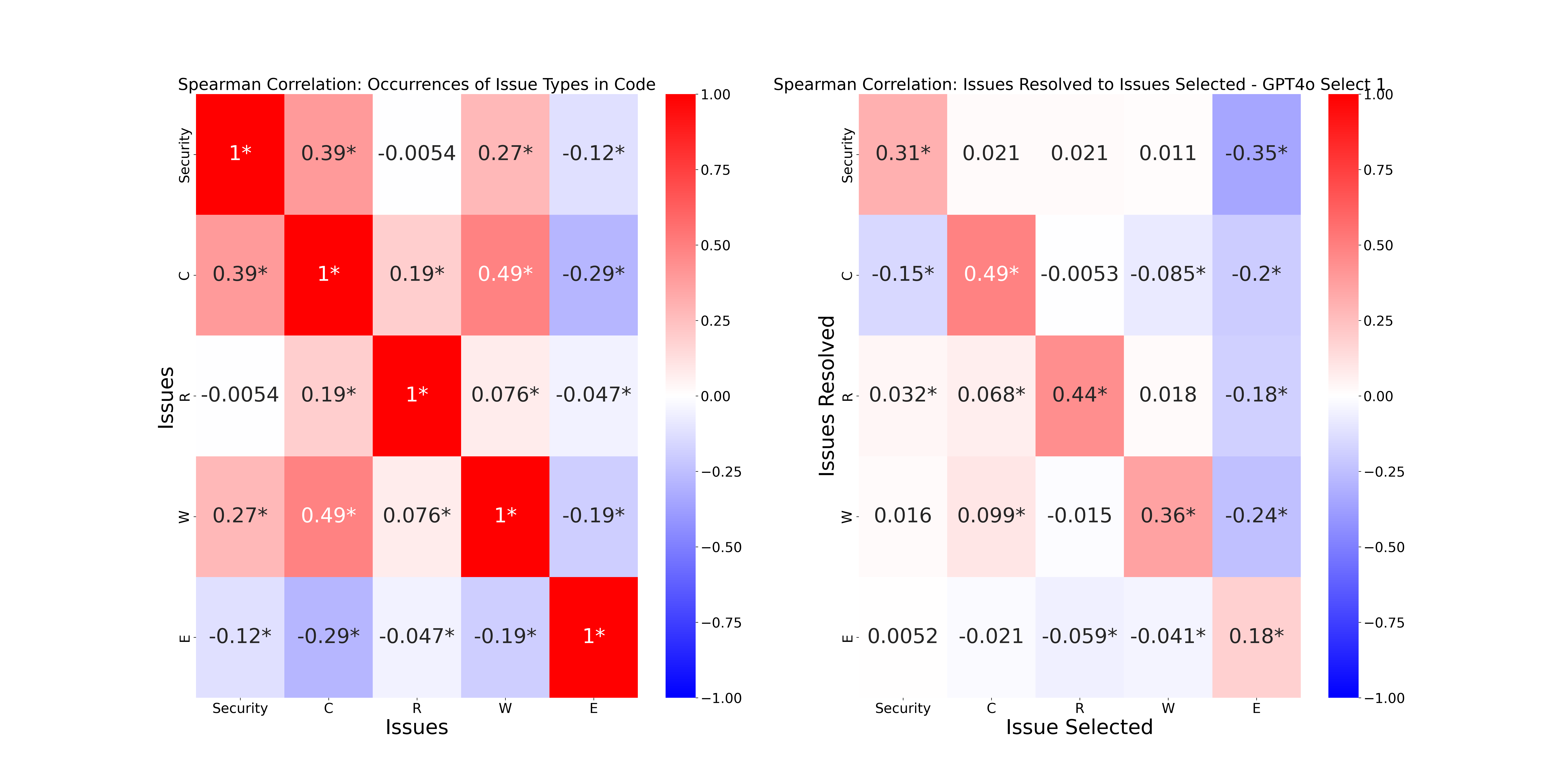}
    \label{fig:GPT4oSpearman}
\end{figure*}

Figure~\ref{fig:GPT4oSpearman} presents Spearman correlation coefficients for the \textit{Select 1} experiment, chosen to control for co-selection effects (e.g., fixing both security and convention issues simultaneously). Key findings include that convention (readability) issues are often introduced, and that fixing error-related issues can introduce new problems.

\begin{table*}
    \centering\setlength{\tabcolsep}{1pt}
    \caption{Top 3 issue types most frequently introduced during refinement.}
    \label{tab:introducingIssues}
    \begin{tabular}{llll} \toprule
        Issue Code & Avg. Change & Times Introduced & Example message (rule name) \\ \midrule
    C0103 & -0.00 & 245 & Variable name \textless variable\textgreater ~doesn't conform to  naming style (invalid-name) \\ 
        C0303 & -0.08 & 217 & Trailing whitespace (trailing-whitespace) \\ 
        C0301 & -0.05 & 149 & Line too long (line-too-long) \\ 
        R1705 & -0.00 & 72 & Unnecessary ``elif'' after ``return'' remove the leading ``el'' from ``elif'' (no-else-return) \\ 
        R1710 & -0.00 & 25 & Either all return statements in a function should return an expression or none (inconsistent-return-statements)\\ 
        R0913 & -0.00 & 14 & Too many arguments (too-many-arguments) \\ 
        W0621 & -0.00 & 140 & Redefining name \textless name\textgreater~from outer scope line \textless line number\textgreater~(redefined-outer-name) \\ 
        W0718 & -0.00 & 131 & Catching too general exception Exception (broad-exception-caught) \\ 
        W1514 & -0.00 & 70 & Using open without explicitly specifying an encoding (unspecified-encoding) \\ 
        E0001 & \hspace{1mm}0.00 & 189 & Syntax error (syntax-error) \\ 
        E0401 & -0.01 & 104 & Unable to import \textless name\textgreater~(import-error) \\ 
        E1101 & -0.01 & 22 & Module \textless moduleName\textgreater~has no \textless memberName\textgreater~member (no-member) \\ \bottomrule
    \end{tabular}
\end{table*}

\begin{table*}
    \centering
    \caption{Issue introduction tendencies across quality categories. Numbers rounded to two decimals.}
    \label{tab:qualityDecrease}
    \begin{tabular}{llll} \toprule
    Category & Avg. Change & Count & Proportion of iterations introducing issues \\ \midrule
    Security & -0.10 & 74 & 0.05 \\ 
    Readability (Convention) & -0.02 & 995 & 0.01 \\ 
    Maintainability (Refactor) & -0.00 & 151 & 0.00 \\ 
    Reliability (Warning) & -0.00 & 529 & 0.00 \\ 
    Functionality (Error) & -0.00 & 337 & 0.00 \\ 
    \bottomrule
    \end{tabular}
\end{table*}

Tables~\ref{tab:introducingIssues} and~\ref{tab:qualityDecrease} quantify the tendency of different issue types to be introduced during refinement. Security issues are disproportionately introduced when attempting to fix code.


\section{Discussion}\label{sec:discussion}

\subsection{RQ1: To what extent can Large Language Models resolve code quality issues?}

\textbf{RQ1.1: What types of issues can be addressed?}

\begin{table*}[]
    \centering
    \caption{Top 10 issues with highest resolution rate across all experiments. 
    Only issues appearing more than 20 times are shown.}
    \label{tab:top10Best}
    \begin{tabular}{lll}
    \toprule
         Code & Resolution Rate & Description \\ \midrule
        W3101 & 1.00 & Missing timeout argument for method \textless method\textgreater, can cause your program to hang indefinitely (missing-timeout) \\
        W0611 & 1.00 & Unused abort imported from \textless library\textgreater~(unused-import) \\
        W0612 & 1.00 & Redefining name \textless name\textgreater~from outer scope line \textless line number\textgreater~(redefined-outer-name) \\ 
        R1705 & 0.93 & Unnecessary ``elif'' after ``return'' remove the leading ``el'' from ``elif'' (no-else-return) \\ 
        W1514 & 0.91 & Using open without explicitly specifying an encoding (unspecified-encoding) \\ 
        C0116 & 0.89 & Missing function or method docstring (missing-function-docstring) \\
        C0115 & 0.89 & Missing class docstring (missing-class-docstring) \\
        R1710 & 0.86 & Either all return statements in a function should return an expression or none (inconsistent-return-statements)\\ 
        C0301 & 0.84 & Line too long (line-too-long) \\ 
        C0114 & 0.83 & Missing module docstring (missing-module-docstring) \\
    \bottomrule
    \end{tabular}
\end{table*}

Although overall code quality has improved, some issues persist, especially within the \textit{Error} category. This raises the question of which issues LLMs are particularly adept at resolving. Table~\ref{tab:top10Best} highlights the most successfully resolved issues.

Notably, LLMs demonstrate strong capabilities in generating documentation in the correct format. For example, C0114 (missing-module docstring), C0116 (missing-function-docstring), and C0115 (missing-class-docstring) show resolution rates of 83\%, 89\%, and 89\%, respectively.

Furthermore, GPT-4o effectively addresses refactoring issues, with two of the top 10 solved issues belonging to this category. Figure~\ref{fig:beforeAfter} corroborates this success, indicating that LLM-driven static analysis can substantially enhance code maintainability.

\textbf{RQ1.2: How much prompting is required to fix these defects?}

A key aspect to consider is when does a code snippet becomes difficult or impossible to improve using LLMs? Additionally, we examine how the hyperparameter \textit{IssuesSelected} influences a LLM's performance.

Table~\ref{tab:performance} indicates that selecting only one issue per iteration ($IssuesSelected = 1$) yields the poorest results across nearly all quality aspects---except for functional correctness. This is likely due to the limited opportunity for iterative refinements. Figure~\ref{fig:probabilityImprovement} shows that \textit{Select 1} retains a ~20\% probability of improving a snippet after 10 iterations, implying that further iterations may be beneficial.

Figure~\ref{fig:gpt4oSeverity} suggests that selecting more issues per iteration enhances LLMs' performance. Table~\ref{tab:issuesSelectedInit} confirms this trend: as \textit{IssuesSelected} increases, the correlation between initial and post-improvement issue counts weakens. This suggests that increasing the number of selected issues enables broader improvements.

In summary, while no single optimal configuration exists, selecting 3, 4, or 5 issues per iteration (with 10 iterations) performs best. In contrast, selecting only 1 or 2 issues per iteration results in incomplete improvements.

\begin{table}[]
    \centering
    \caption{Spearman correlations between initial and post-improvement issue counts for each \textit{IssuesSelected} configuration. }
    \label{tab:issuesSelectedInit}
    \begin{tabular}{lcc}
    \toprule
        IssuesSelected & Spearman Correlation & p-value \\ \midrule
        1 & 0.32 & 0.0000 \\
        2 & 0.14 & 0.0026 \\
        3 & 0.10 & 0.0406 \\
        5 & 0.16 & 0.0005 \\
        All & 0.11 & 0.0235 \\
    \bottomrule
    \end{tabular}
\end{table}

\textbf{RQ1.3: How consistent is the improvement across different quality aspects?}

Our results demonstrate that LLMs consistently resolve static analysis-identified issues across all quality dimensions. Figure~\ref{fig:beforeAfter} shows that issue frequencies decrease across every aspect post-improvement.

For security, the percentage of insecure code snippets drops to 13.41\%--15.32\% after refinement. While this represents a notable success, it does not surpass the 9.4\% vulnerability rate achieved by Alrashedy et al.~\cite{alrashedy2023can}.

Readability improvements are particularly pronounced, with convention issues decreasing from 84.21\%--86.99\% to 11.38\%--32.6\%. Similarly, warning-related issues decline from 54.61\%--59.08\% to 10.75\%--21.66\%.

Despite the substantial improvements across most dimensions, functionality-related issues remain relatively persistent, showing only a 5\%--7\% decrease. As Table~\ref{tab:correctness} illustrates, the proportion of functionally correct snippets increases by at most 10.94\%.

Overall, while the GPT-4o demonstrates clear success in improving code quality, functionality-related improvements remain comparatively limited.

\textbf{RQ2: What new challenges emerge when LLMs attempt to rectify code quality issues?}

Ideally, issues should be resolved without introducing new defects. However, some unintended problems emerge during refinement.

Convention-related issues are the most commonly introduced, with 994 occurrences post-improvement. However, security issues are disproportionately introduced relative to other categories---5\% of iterations result in new security vulnerabilities (Table~\ref{tab:qualityDecrease}).

Exploring correlations among issue introductions reveals interesting patterns. Figure~\ref{fig:GPT4oSpearman}b, based on the \textit{Select 1} experiment, shows relationships between issue resolution and introduction:
\begin{itemize}
    \item Selecting security issues correlates with a slight reduction (-0.15) in convention-related problems.
\item When convention issues are not directly targeted, their frequency either remains unchanged or decreases slightly.
\item Error-related issues negatively correlate with other issue categories, suggesting that resolving functionality problems may introduce security concerns.
\end{itemize}

Despite these tendencies, the comparison of initial and post-improvement snippets indicates that only three unique issues became more prevalent: E1101 (no-member), R0912 (too-many-branches), and F0002 (astroid-error).

\subsection{Summary}

While LLMs significantly improve code quality across multiple aspects, some functional correctness issues remain stubborn. The iterative approach demonstrates effectiveness, though some unintended issues arise during refinement. Convention and security-related problems are most commonly introduced, highlighting areas for future optimization.

Overall, while the tool effectively enhances code readability, maintainability, and reliability, functional correctness and security require additional refinement for further improvements.


\section{Conclusion}\label{sec:conclusion}

As LLMs continue to advance in their code generation capabilities, evaluating their effectiveness beyond functional correctness becomes increasingly important. This study explores existing approaches for improving the security of generated code and methods for assessing code quality comprehensively. After identifying limitations in previous research, we propose an algorithm utilizing static analysis tools---Bandit and Pylint---to evaluate the quality of LLM-generated code across four key dimensions: readability, maintainability, functionality, and security. Additionally, we employ test suites to assess functional correctness.

Building upon this quality metric and leveraging insights from static analysis reports, we developed the \textit{SelectIssues} prompting strategy. \textit{SelectIssues} enhances code quality by injecting issue-specific information directly into the code snippet and prompting the LLM to resolve highlighted problems. The LLM iteratively proposes solutions, accepting only those that improve the code. This iterative refinement process is capped at 10 iterations.

Applying this methodology to the PythonSecurityEval dataset~\cite{alrashedy2023can}, we observed that LLMs were largely successful at resolving issues flagged by static analysis. Improvements were consistently achieved across all investigated quality dimensions: 
\begin{itemize}
    \item Security: Reduction of insecure code by up to 32\%.
\item Readability: Convention-related issues decreased by up to 74\%.
\item Reliability: Warning-related issues reduced by up to 44\%.
\item Functionality: Error-related issues declined by up to 7\%.
\item Correctness: Functional correctness increased by up to 11\%.
\end{itemize}

An important observation was the diminishing returns of iterative improvements: by the third or fourth iteration, many code snippets became resistant to further refinement, and by the 10th iteration, the probability of additional improvement fell below 10\%. Additionally, unintended readability issues 
were frequently introduced. Security concerns were also occasionally exacerbated, with 5\% of proposed solutions containing more security flaws than the original snippet.

\section{Limitations and Future Work}\label{sec:limit}

Despite the effectiveness of our approach, three limitations must be acknowledged. 

\textit{False Positives and False Negatives in Static Analysis.} 
   Static analysis tools occasionally misidentify issues, particularly within the security domain, where false positives are common. Conversely, false negatives---cases where tools fail to detect a genuine code quality issue---also present challenges. However, prior research~\cite{vassallo2020developers} indicates that developers actively use these tools to identify and mitigate quality concerns, reinforcing their value despite imperfections.

\textit{Test Suite Overfitting.} 
   While test suites serve as a useful correctness benchmark, passing all test cases does not necessarily indicate true functional correctness. This overfitting issue is well-documented in Automated Program Repair research~\cite{10.1145/3092703.3092718,petke2023program}. Future work should explore additional correctness validation methods.

\textit{Potential Misalignment in Quality Aspect Mapping.} 
   Our study assigns static analysis issue categories to specific quality dimensions based on existing literature and ISO 25010 standards. However, categorization is not always clear-cut; for example, not all warnings necessarily correlate with reliability (e.g., W0641: ``possibly unused variable''). Further refinement of these mappings could improve classification accuracy.

Future research should focus on enhancing prompt engineering techniques to further mitigate unintended issue introduction, investigating alternative methods to complement static analysis for a more holistic quality assessment, and exploring ways to improve functional correctness beyond test suite verification.

Despite these limitations, our framework demonstrates that LLMs, with structured prompting and iterative refinement, can significantly enhance code quality across multiple dimensions. This study lays the groundwork for further advancements in AI-assisted code generation and optimization.

\section*{Acknowledgment}
Sherlock Licorish and Markus Wagner are funded by the New Zealand Ministry of Business, Innovation and Employment (MBIE) Smart Idea Award UOO2456.

\bibliographystyle{IEEEtran}
\bibliography{sample-base}


\end{document}